\documentstyle[12pt,psfig]{article}
\begin{document}

\title{Inflation from an Effective Standard Model of Particle 
Physics for curved space-time}
\author{Tonatiuh Matos, Guillermo Arreaga \\
%EndAName
{\it Departamento de F\'{\i}sica}\\
{\it Centro de Investigaci\'on y de Estudios Avanzados del IPN}\\
{\it Apdo. Postal 14-740,07000 M\'exico, D.F., MEXICO}\\ \\
\and
Gabriella Piccinelli \\
%EndAName
{\it Instituto de Astronom\'{\i}a}\\
{\it Universidad Nacional Aut\'onoma de M\'exico}\\
{\it Apdo. Postal 70-543, 04510 M\'exico, D.F., MEXICO}}
\maketitle

\begin{abstract}

Beginning from an effective theory in eight dimensions, in Ref.\cite{mcm},
Macias, Camacho and Matos proposed an effective model for the 
electroweak part of the Standard Model of particles in curved 
spacetime. Using this model, we investigate the cosmological 
consequences of the electroweak
interaction in the early universe. We use the approximation that, near the
Planck epoch, the Yang-Mills fields behave like a perfect fluid. Then we
recover the field equations of inflationary cosmology, with the Higgs field
directly related to the inflaton. We present some qualitative discussion
about this and analyse the behavior of isospin space using some
known exact solutions.
\end{abstract}

\newpage

\section{Introduction}
\label{int}

Higher-dimensional theories are one of the most old and
interesting unification theories considered by physicist and mathematicians
unifying gravitation with Yang-Mills interactions. The history of these
theories began with Theodore Kaluza and Oscar Klein, when they formulated a
five-dimensional theory unifying gravitation and electromagnetism, but this
theory used a great amount of hypothesis on the structure of space-time
and on the functional dependence of physical quantities. The first real
geometrical formulation of these theories was given by 
Kerner \cite{kerner} and Cho \cite{cho} (see also \cite{ma}). They 
formulated the theory as a principal fiber bundle, being the 
four-dimensional space-time the basis space of the bundle, and 
the gauge group of the Yang-Mills fields they wanted to unify with 
gravitation, the typical fiber. This
formulation pretended to be a fundamental theory of all interactions,
because it unified all the Yang-Mills gauge theories with gravitation.
Nevertheless, it is now believed that a superstrings theory could be the
fundamental theory of all interactions, among other reasons, because it
contains the Kaluza-Klein theories (KK) in its low energy 
limit. In this case,
KK become effective theories for Yang-Mills and gravitational fields 
instead of fundamental ones.

In Ref.\cite{mcm}, Macias, Camacho and Matos (MCM) considered
an effective eight-dimensional theory, supplemented with the Yukawa 
and the fermionic sectors as in the Standard Model and, as 
expected, the Weinberg, Glashow and Salam Model was recovered (see also \cite
{knr} and \cite{fman}). Then, some questions arose. How is the cosmology
predicted by this Standard Model in curved space-time? Is there inflation in
this cosmology? Which scalar field plays the role of the inflaton? What 
is its relation with the Higgs field? In this paper we begin with the 
eight-dimensional
Kaluza-Klein-Dirac theory of Ref.\cite{mcm} in order construct the cosmology
of an effective
particle theory for the electroweak Standard Model in a curved
space-time, and answer at least partially these
questions. 

Following Kerner \cite{kerner} and Cho \cite{cho}, who
established the appropiated general structure of the metric on the bundle in
order to consider a non-Abelian Lie gauge Group $G$, MCM
set up the metric:

\begin{equation}
\begin{array}{ll}
ds^{2} & =g_{\hat{\mu}\hat{\nu}}\;dx^{\hat{\mu}}\otimes dx^{\hat{\nu}%
}=g_{\mu \nu }\;dx^{\mu }\otimes dx^{\nu }\vspace{0.3cm} \\ 
& \hspace{-1cm}-I_{1}^{2}(x)\left[ dx^{5}+k\;A_{\mu }(x)\;dx^{\mu }\right]
\otimes \left[ dx^{5}+k\;A_{\nu }(x)\;dx^{\nu }\right] \vspace{0.3cm} \\ 
& \hspace{-1cm}-I^{\dagger }(x)I(x)\;\gamma _{ij}(y)\left[ dy^{i}+\frac{k}{%
L_{.}}\;K_{\alpha }^{i}(y)\;A_{\mu }^{\alpha }(x)\;dx^{\mu }\right] \otimes 
\left[ dy^{j}+\frac{k}{L_{.}}\;K_{\beta }^{j}(y)\;A_{\nu }^{\beta
}(x)\;dx^{\nu }\right] \;,
\end{array}
\label{metrica8dim}
\end{equation}
where $\gamma_{i j}(y)$ is the metric of the isospace manifold, 
$K_{\alpha}^i(y)$ are its Killing vectors, $A_{\mu}(x)$ and 
$A_{\mu}^{\alpha}(x)$
are the gauge potentials.  MCM considered 
$SU(2)\times U(1)$ as a structure group of the 
eletroweak interactions. Here $I_{1}(x)$ plays the role of the dilaton of the
theory and also gives the radius of the $S^{1}$ part of the isospace. The
field $I(x)$ is endowed with a three dimensional spinorial structure and
plays the role of a Higgs field in the effective theory. This field enters
into the metric through the scalar combination $I^{\dagger }(x)I(x)$ and is
associated to the radius of the $S^{3}$ part of the isospace.
In Eq.(\ref{metrica8dim}), $k$ and $L$ are scaling costants.

They used the following action

\begin{equation}
I_{8}=\int d^{4}xd^{4}y\sqrt{-\hat{g}}\left( \frac{1}{16\pi GV}\left[ \hat{R}%
+V+Y_{u}\right] +{\cal L}_{D}\right) \;.  \label{acd8}
\end{equation}
In Eq.(\ref{acd8}), $V$ is the Higgs potential and $Y_{u}$ represents the terms
for the Yukawa couplings. From this theory, they obtained the four-dimensional
effective eletroweak part of the Standard Model. It was shown in 
\cite{mcm} that the 
masses of the gauge bosons and of the first two fermion families were correctly 
given by this theory. Here, we are interested in studying the
cosmological implications of this kind of approach. We can represent the matter content
of the universe as a perfect fluid describing a highly energetic and isotropic
radiation. Then, instead of the Dirac Lagrangian, ${\cal L}_{D}$, we will
use a perfect fluid Lagrangian. We will also neglect the Yukawa 
interactions terms $Y_{\mu}$.

Finally,
we will assume that the metric $g_{\mu \nu }$ for the base 
space-time ${\cal M}$ 
is given by 
the Robertson-Walker metric and that both the scalar $I_{1}(x)$ and spinorial 
$I(x)$ fields depend only on time, $x^{0}=t$; otherwise, they could be
used to distinguish between different directions in space-time, invalidating 
our assumption of the isotropy of the early universe. 
In this way, all of the terms involving the
gauge fields, $A_{\mu }$, $A_{\mu }^{\alpha }$, do not have to be explicitely
accounted for in writing the final equations of motion.

From this model we obtain that the inflaton $\sigma $ is produced in a natural 
way as a combination of the scalar fields $I_{1}(t)$ 
and $I^{\dagger}(t)I(t)$; in this way, we asign a geometrical origin 
to the inflaton: it is
related to the radius of the internal space, the isospin space. On the other
hand, inflation takes place by a phase transition of the
(electroweak) Higgs field, i.e. the scalar field that generates inflation and
the scalar field that gives mass to particles are related. So, depending on
the particular inflationary potential that is used, this model allows 
that these two roles, of the same field, may be played in two 
different steps. According to this model, a part of the internal 
space deflates while the space-time inflates; after some time, the 
scalar field settles to a constant, while the space-time keeps 
expanding. Then another question arises: how can the same scalar 
field be responsible for these two different phenomena at the 
origin of the Universe? In the concluding remarks we give a 
partial answer to this problem. Using the simplification of 
modelling the matter contribution as a perfect fluid, we 
obtain an approximation of the behavior of the internal space 
and the well-known
evolution of space-time.

\section{Effective four--dimensional action}
\label{dimred}

We start with metric (\ref{metrica8dim}). Following MCM, ( see 
also J. Toms 
in his review essay on the Kaluza--Klein theory, \cite{toms} ) in order 
to  easily calculate geometrical quantities of this metric, we change 
to a horizontal lift basis $\hat{e}_{%
\hat{\mu}}$ defined by the relations

\begin{equation}  \label{base}
\hat{e}_{\hat{\mu}}\equiv \left\{ 
\begin{array}{l}
e_{\mu}=\partial_{\mu}-k A_{\mu}\partial_5 - \frac{k}{L}\;\;
A^{\alpha}_{\mu} K^{i}_{\alpha}\partial_{i}\vspace{0.3 cm} \\ 
e_5 = \partial_5 \vspace{0.3 cm} \\ 
e_{i}= \partial_{i} .
\end{array}
\right.
\end{equation}
Evidently, this basis is not holonomic; its commutation relations 
can be easily
derived (see T. Matos and A. Nieto, \cite{ma}). One finds

\begin{equation}  \label{conmutador}
\begin{array}{ll}
\left[ e_{\mu}, e_{\nu}\right] & = - k\; F_{\mu \nu}\; e_5 -\frac{k}{L}\;
K^i_{\alpha}\; F^{\alpha}_{\mu \nu}\; e_i \vspace{0.3 cm} \\ 
\left[ e_{\mu}, e_i\right] & = \frac{k}{L}\; A^{\alpha}_{\mu}\; \left(
\partial_i K^j_{\alpha}\right) \; e_j \vspace{0.3 cm} \\ 
\left[ e_{\mu}, e_5 \right] & = 0,
\end{array}
\end{equation}

\noindent where

\begin{equation}  \label{camposdef8dim}
\begin{array}{ll}
F_{\mu \nu} & =\partial_{\mu}A_{\nu}-\partial_{\nu}A_{\mu}\vspace{0.3 cm} \\ 
F^{\alpha}_{\mu \nu} & = \partial_{\mu}A^{\alpha}_{\nu}-
\partial_{\nu}A^{\alpha}_{\mu} +\frac{k}{L}\; f_{\beta \gamma}^{\; \alpha}
A^{\beta}_{\mu}\; A^{\gamma}_{\nu}
\end{array}
\end{equation}

\noindent are the field strengths for the potentials $A_{\mu}$ and 
$A^{\alpha}_{\mu}$, respectively. In this basis, the metric takes a
diagonal form. Finally, the space Ricci scalar curvature reads

\begin{equation}  \label{curvatura8dim}
\begin{array}{ll}
\hat{R}= & R-\frac{1}{I^{\dagger}I}\tilde{R} -\frac{1}{4} \left( \frac{k}{L}
\right)^2 I^{\dagger}I \gamma_{i j} K^i_{\alpha} K^j_{\beta} F^{\alpha \mu
\nu} F^{\beta}_{\mu \nu}-\frac{1}{4} k^2 I_1^2 F^{\mu \nu} F_{\mu \nu}%
\vspace{0.3 cm} \\ 
& +3\left(\partial^{\mu} \log I^{\dagger}I\right)_{;\mu}+3\partial_{\mu}
\log I^{\dagger}I \partial^{\mu}\log I^{\dagger}I\vspace{0.3 cm} \\ 
& + 2 \; \left(\partial^{\mu}\log I_1\right)_{;\mu}+ 2 \;\partial_{\mu} \log
I_1 \partial^{\mu} \log I_1 + 3 \; \partial_{\mu} \log
I^{\dagger}I\partial^{\mu} \log I_1\vspace{0.3 cm} \\ 
& +\frac{3}{2}g^{\mu \nu}_{\hspace{0.3 cm},\nu} \partial_{\mu} \log
I^{\dagger}I +g^{\mu \nu}_{\hspace{0.3 cm},\nu} \partial_{\mu} \log I_1,
\end{array}
\end{equation}

\noindent where $R$ is the scalar curvature of the space-time manifold 
${\cal M}$ and $\tilde{R}$, which is a constant in our case, is the 
scalar curvature of
the isospace manifold $S^1\times S^3$.

It should be noticed that we have used an effective version of 
Kaluza-Klein theory. It seems to us that this is the most appropiate way to
include the Standard Model into a theory of gravitation, see Refs.\cite{knr}
and \cite{fman}.

Let us consider the following action defined in the eight--dimensional
double fiber bundle space ${\cal {P}} \sim {\cal M}\times SU(2) \times U(1)$:

\begin{equation}
{\cal S}=\int d^4 x d^4 y \; \sqrt{-\hat{g}}\;
\frac{1}{16\pi GVL_{1}}\left( \hat{R}+
\hat{V}(I_{1},I)+\hat{{\cal L}}\right),  
\label{at8d}
\end{equation}

\noindent where we introduce $L_1$ on dimensional grounds.
We begin using the eight-dimensional theory with a potential of the form

\begin{equation}  
\label{potform}
\hat{V}\left( I_1,I \right)\equiv 
\hat{V}_1 \left( \sqrt{\frac{3}{8 \pi G}}\; \log
I^{\dagger}I \right) + 
\hat{V}_2 \left( -\sqrt{\frac{1}{6 \pi G}}\; \log I_1 \right)
\end{equation}

\noindent and we include the lagrangian $\hat{{\cal {L}}}$ to represent, when
the dimensional compactification has been achieved, the matter content of the
universe.

According to Eq.(\ref{metrica8dim}), the determinant of the
metric, $\hat{g}$, is given by

\begin{equation}  \label{det8dim}
\sqrt{-\hat{g}}= \sqrt{-g}\sqrt{\gamma}\; I_1 \left( I^{\dagger}
I\right)^{3/2},
\end{equation}

\noindent where $g\equiv \det(g_{\mu \nu})$ and $\gamma \equiv
\det(\gamma_{i j})$. The coupling between the curvature scalar and 
the fields $I_1(x)$ and $I^{\dagger}I(x)$ will lead us, after 
dimensional reduction, to a Brans--Dicke theory. We prefer to 
make a conformal transformation
in order to avoid this coupling.

Let $f(x)$ be a function which depends only on space-time coordinates of 
${\cal {M}}$. We multiply the internal part of metric $\hat{g}_{\hat{\mu} 
\hat{\nu}}$, Eq. (\ref{metrica8dim}), by the function $f^2(x)$, so we can
define new scale functions for the $S^1\times S^3$ group manifold, by
means of

\begin{equation}  \label{newfields}
K(x)\equiv f(x)I_1(x)\hspace{1 cm}\& \hspace{1 cm} F(x)\equiv f(x)I(x) .
\end{equation}

We can obtain the Ricci curvature associated with this conformal transformed
metric $\hat{g}^{\prime}_{\hat{\mu} \hat{\nu}}$ by 
substituting $I_1\rightarrow K$ and $I\rightarrow F$ in the 
expression previously found for Ricci scalar curvature, 
Eq.(\ref{curvatura8dim}).

If we set $f=\left[I_1 (I^{\dagger}I)^{3/2} \right]^{-\frac{1}{4}}$, 
we obtain that the determinat $\hat{g}^{\prime}$ is
given by the equation

\begin{equation}  \label{detsep8dim}
\sqrt{-\hat{g}^{\prime}}= \sqrt{-g}\sqrt{\gamma}.
\end{equation}

The relations between the old and the new fields are:

\begin{equation}  \label{nuevasidentidades}
\begin{array}{c}
\log F^{\dagger}F = -\frac{1}{2} \log I_1 + \frac{1}{4} \log I^{\dagger}I 
\vspace{0.3 cm} \\ 
\log K=\frac{3}{4}\log I_1 - \frac{3}{8} \log I^{\dagger}I.
\end{array}
\end{equation}

Let us now introduce the following useful definitions for the 
fields $\Phi$ and $\Psi$:

\begin{equation}  
\label{caray}
\begin{array}{l}
\sqrt{\frac{3}{16}}\log I^{\dagger}I= \frac{1}{\sqrt{2}}\Phi
\vspace{0.3 cm}\\ 
\sqrt{\frac{3}{4}}\log I_1=\frac{1}{\sqrt{2}}\Psi .
\end{array}
\end{equation}

\noindent We can further make the identification of
the scalar field, $\sigma$, with a combination of the fields previously
defined, and two additional definitions for an effective cosmological
constant and for the dimensional factor $k^2$:

\begin{equation}  
\label{sus4dimsegprim}
\begin{array}{l}
\sigma = \frac{\Phi -\Psi}{\sqrt{16 \pi G}}\vspace{0.3 cm} , \\ 
\hat{\Lambda}=\frac{1}{2} \tilde{R} \exp \left(-\sqrt{\frac{8 \pi G}{3}}\;
\sigma\right) , \vspace{0.3 cm} \\ 
k^2\rightarrow 16 \pi G.
\end{array}
\end{equation}

We finally get an effective four dimensional space-time action, that is

\begin{equation}  
\label{actionefec}
\begin{array}{ll}
{\cal A} & = \int d^4x \sqrt{-g}\;\Bigg\{ \frac{1}{16\pi G}\left[ R-2\hat{%
\Lambda} \right] \vspace{0.3 cm} \\ 
& -\frac{1}{4} \exp \left(-\sqrt{24 \pi G}\;\sigma \right)F_{\mu \nu}F^{\mu
\nu} -\frac{1}{4} \exp \left(\sqrt{\frac{8\pi G}{3}}\;\sigma \right)
F^{\alpha}_{\mu \nu}F^{\alpha \mu \nu}\vspace{0.3 cm} \\ 
& +\frac{1}{2} \partial_{\mu}\sigma\partial^{\mu}\sigma -V(\sigma) +{\cal L}%
_{matter}\Bigg\}.
\end{array}
\end{equation}

\noindent Then, we are left with an Einstein-Hilbert--Yang-Mills effective
action, which presents an interaction between the scalar field and the 
Yang-Mills fields. This action is the main result of this section. 

It should be noticed that two surface terms
have been dropped out in obtaining the effective action (\ref{actionefec}). 
This can actually be done by imposing the following boundary conditions

\begin{equation}
\label{front}
\begin{array}{l}
\lim _{x\rightarrow \infty} \partial^{\mu} \log K \rightarrow 0
\vspace{0.3 cm}\\
\lim _{x\rightarrow \infty} \partial^{\mu} \log F^{\dagger} F \rightarrow 0.
\end{array} 
\end{equation}
Summarizing, the process we have followed to get an effective four 
dimensional action ${\cal A}$ from
the eight--dimensional action ${\cal S}$, is 

\begin{equation}
\label{procred}
{\cal S}\stackrel{conf. trans.}{\rightarrow} {\cal S}'
\stackrel{integration}{\rightarrow} {\cal A}
\end{equation}

\noindent which is usually called a dimensional compactification
procedure by isometry, see Ref. \cite{cho1}.

\section{Equations of Motion and some examples}
\label{finale}

It is straightforward to get the equation of motion for the different fields
of this theory from the effective four--dimensional action ${\cal A}$, Eq.(%
\ref{actionefec}). The corresponding Einstein equations read

\begin{equation}  \label{moveins}
\begin{array}{ll}
G_{\mu \nu}= & -8\pi G \Bigg\{ \partial_{\mu}\sigma\partial_{\nu}\sigma -%
\frac{1}{2} g_{\mu \nu} \partial_{\rho}\sigma\partial^{\rho}\sigma +g_{\mu
\nu}V_{eff}(\sigma) \vspace{0.3 cm} \\ 
& -\exp \left( -\sqrt{24\pi G}\;\sigma\right)\left(F^{\rho}_{\mu}F_{\nu
\rho} -\frac{1}{4} g_{\mu \nu} F^{\tau \rho}F_{\tau \rho}\right)\vspace{0.3
cm} \\ 
& -\exp \left(\sqrt{\frac{8\pi G}{3}}\;\sigma\right)\left(F^{\alpha
\rho}_{\mu} F^{\alpha}_{\nu \rho}-\frac{1}{4} g_{\mu \nu} F^{\alpha \tau
\rho} F_{\alpha \tau \rho}\right)+ T_{\mu \nu}\Bigg\},
\end{array}
\end{equation}

\noindent where we have made use of the following definition

\begin{equation}  \label{poteefec}
V_{eff}(\sigma)\equiv V(\sigma)+\frac{\hat{\Lambda}(\sigma)}{8 \pi G}
\end{equation}

\noindent and $T_{\mu \nu}$ stands for the energy-momentum tensor of the
matter content of the universe.

\noindent In turn, the equation for the scalar field $\sigma$ becomes

\begin{equation}  \label{sigmamov}
\begin{array}{l}
\partial_{\mu}\partial^{\mu} \sigma +\Gamma_{\mu \lambda}^{\lambda}
\partial^{\mu} \sigma + \frac{\partial V_{eff}}{\partial \sigma} -\frac{1}{4} 
\sqrt{24 \pi G}\; \exp \left(-\sqrt{24 \pi G}\;\sigma \right) F^{\mu
\nu}F_{\mu \nu}\vspace{0.3 cm} \\ 
+\frac{1}{4}\sqrt{\frac{8\pi G}{3}} \exp \left(\sqrt{\frac{8\pi G}{3}}
\sigma\right)F^{\alpha \mu \nu}F_{\alpha \mu \nu}=0.
\end{array}
\end{equation}

\noindent Finally the equations for the Yang-Mills fields are

\begin{equation}  \label{campomov}
\partial_{\nu}F^{\mu \nu} + \Gamma_{\nu \lambda}^{\lambda}F^{\mu \nu} -\sqrt{%
24 \pi G}\; \left( \partial_{\nu}\; \sigma \right) \; F^{\mu \nu}=0
\end{equation}

\noindent and

\begin{equation}  \label{campomov*}
\partial_{\nu}F^{\mu \nu}_{\alpha} + \Gamma_{\nu \lambda}^{\lambda}
F_{\alpha}^{\mu \nu} + \sqrt{\frac{8 \pi G}{3}}\; \left(\partial_{\nu} \;
\sigma \right) F_{\alpha}^{\mu \nu}- \frac{k}{L} \delta_{\gamma \beta}
f^{\gamma}_{\alpha \delta} F^{\beta \mu \sigma}A_{\sigma}^{\delta}=0.
\end{equation}

\noindent There is, additionally, a conservation equation 
$G^{\mu \nu}_{\hspace{0.3 cm};\mu}=0$.

It is interesting to note the role played in this system of equations by the
isospace scale factors $K^2$ and $F^{\dagger}F$ of the conformally transformed
metric $\hat{g}^{\prime}_{\hat{\mu} \hat{\nu}}$. They appear explicitly
through the combination

\begin{equation}  \label{scalefactors}
\begin{array}{l}
F^{\dagger}F= \exp \left( \sqrt{\frac{8 \pi G}{3}}\; \sigma \right) \vspace{%
0.3 cm} \\ 
K^2 = \exp \left( -\sqrt{24 \pi G}\; \sigma \right),
\end{array}
\end{equation}

\noindent which can be deduced at once from equations (\ref
{nuevasidentidades}), (\ref{caray}), and (\ref{sus4dimsegprim}). We see, in
Einstein's equations (\ref{moveins}), that $K^2$ appears as an effective
coupling constant for the electromagnetic field. Analogously, the scale
factor $F^{\dagger}F$ appears in front of the Yang-Mills energy--momentum
tensor density. Both these factors appear again in the scalar field
equation of motion, Eq.(\ref{sigmamov}).

It should be noted that the scale factors $K^2$ and $F^{\dagger}F$ are not
independent but are related through the equation

\begin{equation}  \label{rel}
\left( F^{\dagger}F \right)^3 K^2 =1.
\end{equation}
It is also interesting that the scalar field $\sigma$ should satisfy the
boundary condition

\begin{equation}
\label{boundary}
\lim_{x \rightarrow \infty} \; \partial_{\mu} \sigma \rightarrow 0,
\end{equation}

\noindent which is a consequence of the boundary condition
we imposed on the fields $K^2$ and $F^{\dagger}F$, 
Eq. (\ref{front}). This condition translates in a limitation on the 
asymptotic behavior of the scalar field
$\sigma$ and therefore, on the whole theory we are constructing.

Let us consider a very special case of the system of equations derived from
the effective four--dimesional action ${\cal A}$, Eq. (\ref{actionefec}). If
we consider that the metric $\hat{g}_{\hat{\mu} \hat{\nu}}$, Eq. (\ref
{metrica8dim}), describes the very early universe, then we can assume the
following statements:

\begin{enumerate}
\item  We can represent the matter content of the universe as a perfect
fluid of a highly energetic and isotropic radiation, which presents no
interaction. The energy momentum tensor associated with the lagrangian 
${\cal L}_{matter}$ is then given by 
$T_{\mu \nu}=(\rho + p )u_{\mu}u_{\nu}- pg_{\mu \nu}$.

\item  In this way, all the terms involving the gauge fields, $A_{\mu}$, $%
A^{\alpha}_{\mu}$ do not have to be taken explicitly into account.

\item  It is assumed that the universe was isotropical in its early stages
of evolution. Consequently, the metric $g_{\mu \nu}$ for the base 
space ${\cal M}$ is taken as the Robertson-Walker metric and both the 
scalar $I_1(x)$ and
spinorial $I(x)$ fields are taken to depend only on time. 

\end{enumerate}

In this case, the system of differential equations, 
Eqs.(\ref{moveins})-(\ref{campomov*}) becomes

\begin{equation}  \label{sistfinal}
\begin{array}{c}
3\left( \frac{\dot{R}^2}{R^2} + \frac{k}{R^2}\right) = 8\pi G \left( \rho +%
\frac{1}{2} \dot{\sigma}^2 + V_{eff}(\sigma)\right)\vspace{0.3 cm} \\ 
\ddot{\sigma}+3\frac{\dot{R}}{R}\dot{\sigma}+\frac{d V_{eff}} {%
d \sigma} = 0\vspace{0.3 cm} \\ 
\dot{\rho} + 3\frac{\dot{R}}{R}\left(\rho + p\right) =0\vspace{0.3 cm} \\ 
p =\omega \rho,
\end{array}
\end{equation}

\noindent which is clearly the system of equations for the standard
cosmological model filled with a perfect fluid and endowed with a scalar
field.

The potential term $V_{eff}$ has been taken arbitrarily until now 
but, with an appropriate
choice, this system of equations can lead us to inflationary 
solutions; in this sense, we can argue that the scalar field 
$\sigma$ may play the role
of the inflaton of the very early universe.

There exist well known inflationary solutions for this system of 
equations; see for example Ref. \cite{olive} for a review of the 
inflationary paradigm. Most of 
such inflationay solutions are in agreement with the boundary 
conditions we imposed for the inflaton $\sigma$ 
in Eq.(\ref{boundary}). What this condition means concerning the inflaton 
is that, after a long enough time, say $t_0$, the inflaton 
settles down to its {\it vacuum value} $\sigma_0$.  

Let us follow the evolution of the internal space resulting 
from some specific behavior of the scalar field $\sigma$. We choose 
a couple of exact inflationary solutions for the system of 
equations (\ref{sistfinal}), which were 
found in Ref. \cite{matos} (see also, for example, \cite{bfk} and 
\cite{gk} for a singular-free cosmology). These
solutions are not realistic since they do not describe radiation or matter
dominated epochs, but have the advantage that they are exact and follow the
evolution of the scalar field through several epochs of expansion 
of a multi component universe . The first case corresponds to the potential $
V_{eff}=V(\sigma )=\lambda (\sigma ^{2}+\delta ^{2})^{2}$, with a negative
cosmological constant. The solution describes a closed universe, with a scale
factor given by $a(t)=a_{0}\sin ^{2}(\sqrt{\lambda }\delta t)$. The exact
solution for the scalar field is $\sigma (t)=\delta \cot (\sqrt{\lambda }%
\delta t)$ (see figure 1). When $t$ goes to zero, the scalar field $\sigma
_{1}$ goes to infinity, then the universe inflates and the scalar field
experiences a rapid decay followed by an epoch of slow variation, during the
standard decelarated expansion. The behavior of the isospace scale factors
is: $F^{\dagger}F$ initiates at infinity and $K^{2}$ begins from zero and
eventually both fields remain almost constant (see figure 1). Since 
this solution is periodic, when the spacetime starts to contract again 
the scalar field decreases and goes to $-\infty $ after a deflationary epoch.

Another interesting solution from which we can see the qualitative behavior
of the fields, is given by $\sigma (t)=\delta \coth (\sqrt{\lambda }\delta
t) $; $a_{2}(t)=a_{0}\sinh ^{2}(\sqrt{\lambda }\delta t)$, for the potential 
$V_{eff}(\sigma)=\lambda (\sigma ^{2}-\delta ^{2})^{2}$. Here, the scalar
field goes to infinity as $t$ approaches zero, but it quickly settles in
the constant value $\delta $, which correspondes to the minimum of the
scalar potential $V_{eff}$, i.e. the system experiences a phase transition 
as $t$ separates from zero. Meanwhile, the spacetime inflates exponentially
and the $S^{3}$ part of the innerspace contracts quickly reaching a
constant value, while the $S^{1}$ part of the innerspace grows from zero,
reaches a constant value, and remains there (see figure 2). This
means that a universe with these features experiences an inflationary
spacetime epoch while part of the isospin space contracts and the other part
expands.

\section{Concluding Remarks}

\label{conclu}

We used an effective theory in eight dimensions that contains the
Standard Model, in the sense that, in the weak gravitational field limit, we
can recover all the predictions of the Standard Model. So we can
consider it as an effective model for the Standard Model in curved space and
we consider it is interesting to study the cosmology resulting 
from it. It can be used to study the influence of the electroweak 
interaction on the very early
universe.

We obtain a geometrical interpretation for the scalar field $\sigma $, as a
combination of the innerspace radii. The scale factors that were identified
with Higgs fields in \cite{mcm}, are recombined here in a scalar field that
can be associated to the inflaton. An important issue for this
idendification is that the inflationary evolution leads to a reasonable
behavior for the internal radii. We have proved that, for two particular
exact solutions, the two innerspace radii effectively tend to a constant value
soon after the inflationary regime. In such a way, they are rendered
invisible in the present universe.

We still have to build a complete picture and understand how this field,
that can be identified with a Higgs and an inflaton, manages to
play both roles effectively. We think that the solution should 
be in two stages, i.e.
that each process corresponds to a different step.

Unfortunatly, due to the simplification we are making (namely the
description of the Yang-Mills fields as a perfect fluid) we cannot
quantitatively predict the system's behavior after the potential reaches its
minimum; one expects that quantum fluctuations provoke a reheating mechanism
and maybe the system gets to a local maximum of the potential 
(see \cite{fut}). A standard cosmological phase transition occurs 
by the rolling of the Higgs field from a metastable maximum to a 
minimum of the potential. In our case, the scalar field $\sigma $ 
goes to the local minimum through infinity,
without passing through the metastabe maximum of the potential 
$V_{eff}(\sigma )$. However, we can expect that the scalar 
field $\sigma $ would undergo another phase transition at 
a later stage, giving the adequate masses to fermions.

\vspace{0.5 cm}

{\bf Acknowledgments}

\vspace{0.5 cm}

This work was partially supported by CONACyT M\'exico,
grant 3697-E. G.A. thanks Ulises Nucamendi for useful 
discussions. G.P. acknowledges partial support from DGAPA--UNAM.
We also thank Alfredo Diaz from IA-UNAM for his assistance using \LaTeX.

\newpage

\begin{figure}[tbp]
\label{graf1} \centerline{\psfig{figure=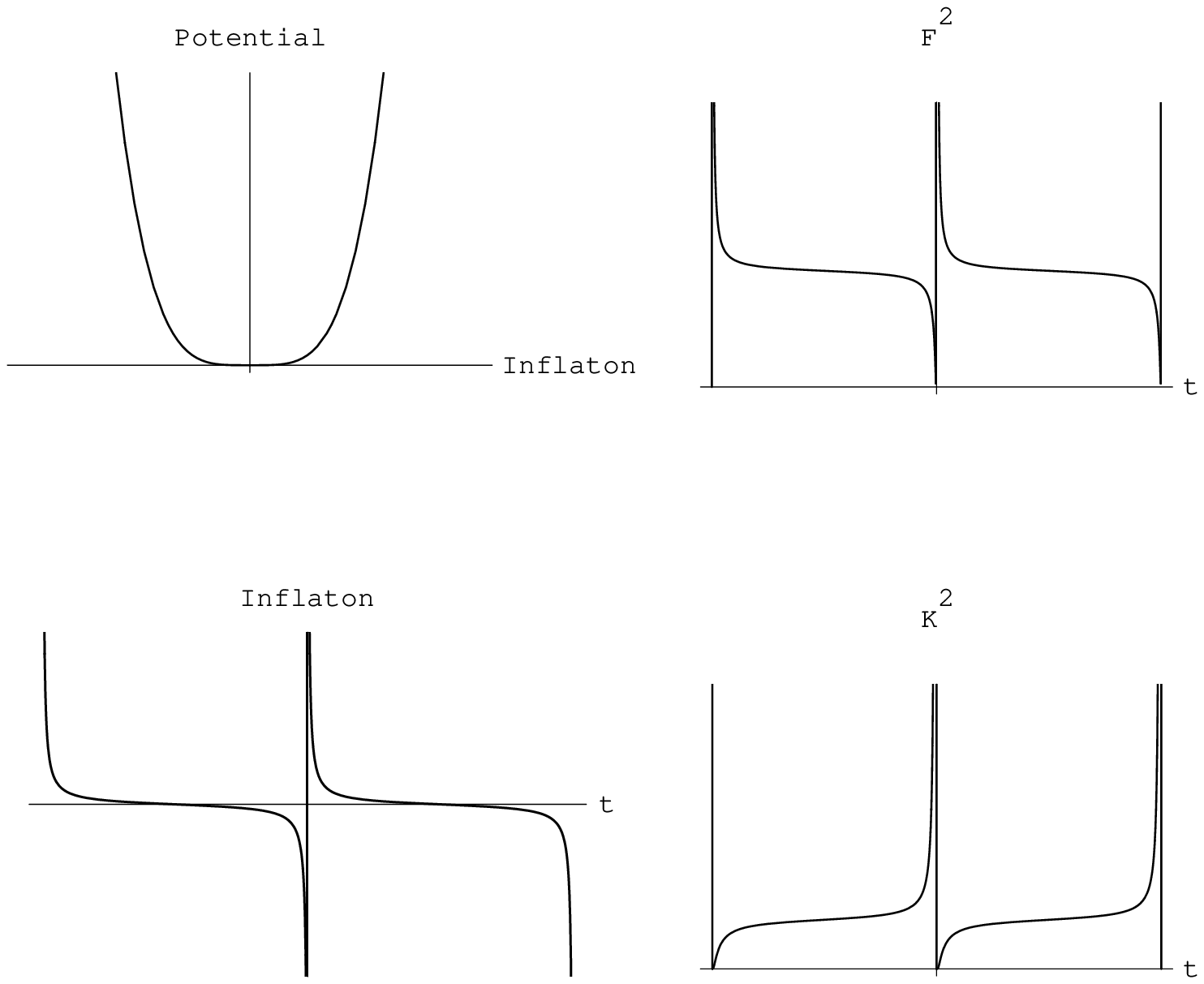}}
\caption{The potential $V(\protect\sigma )= \protect\lambda \left( \protect%
\sigma^2 + \protect\delta^2 \right)^2$ is shown in the top left panel and
the inflaton evolution $\protect\sigma= \protect\delta \cot \left(\protect%
\sqrt{\protect\lambda} \protect\delta t \right)$ in the bottom left. In the
right panels, the scale factors $F^{\dagger}F =\exp \left(\protect\sqrt{%
\frac{8\protect\pi G}{3}} \protect\sigma \right)$ and $K^2=\exp \left(-%
\protect\sqrt{24 \protect\pi G} \protect\sigma \right)$ of the manifolds $%
S^3 $ and $S^1$ are plotted as a function of cosmic time.}
\end{figure}

\newpage

\begin{figure}[tbp]
\label{graf2} \centerline{\psfig{figure=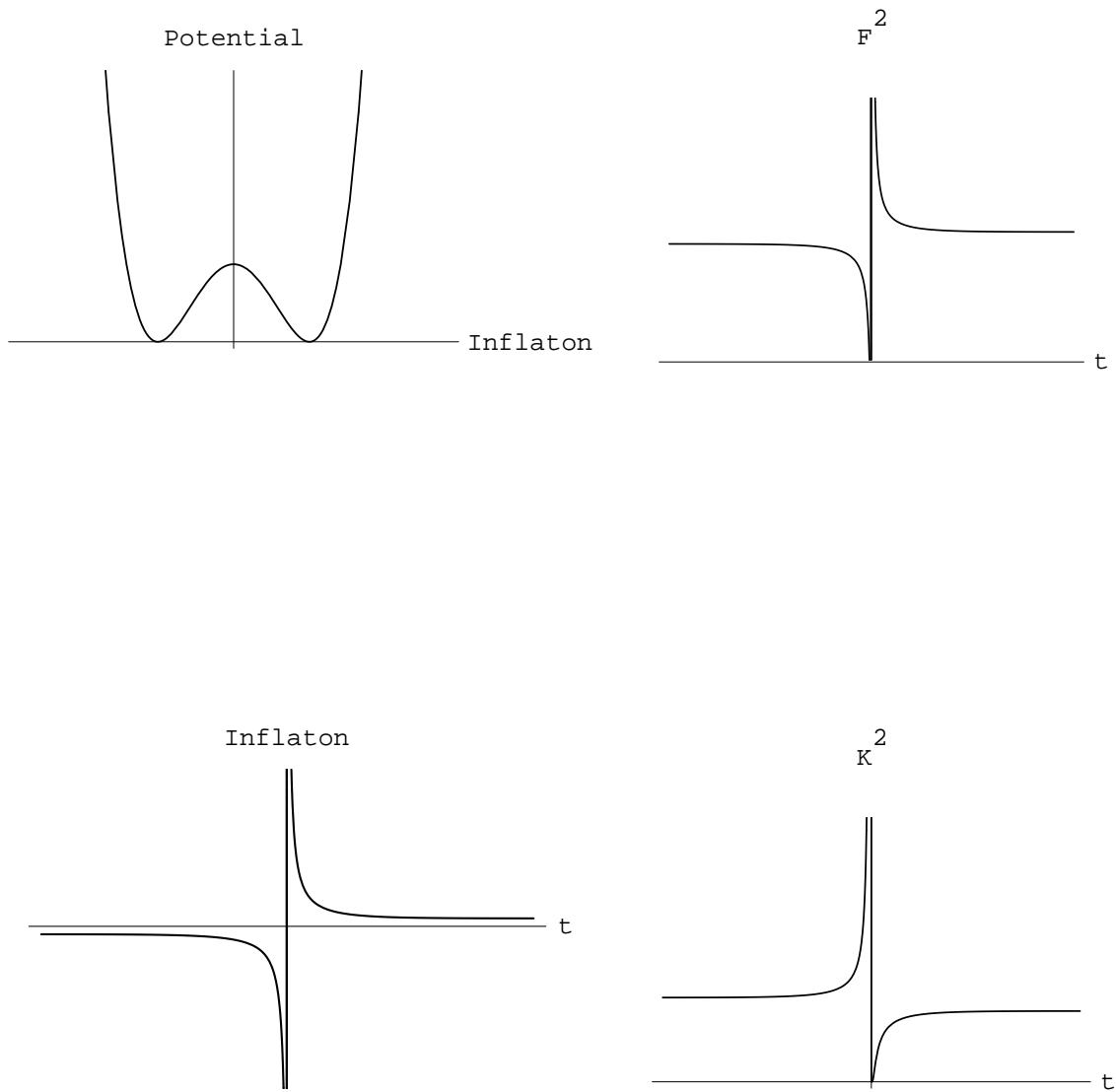}}
\caption{The potential $U(\protect\rho)= \protect\lambda (\protect\rho^2-%
\protect\delta^2)^2$ and the time evolution of the inflaton $\protect\rho = 
\protect\delta \coth \left ( \protect\sqrt{\protect\lambda} \protect\delta t
\right )$ are displayed in the upper and lower left panels, respectively.
The behavior of the scale factors of manifolds $S^3$ and $S^1$, \ \ \ $%
F^{\dagger}F=\exp \left(\protect\sqrt{\frac{8\protect\pi G}{3}} \protect\rho
\right)$ and $K^2=\exp \left(-\protect\sqrt{24 \protect\pi G} \protect\rho
\right)$, is shown in the right panels.}
\end{figure}
\newpage

\end{document}